\documentclass{article} \textheight8.5in \textwidth6.5in
\begin{document}
 \def\ov{\overline} \newtheorem{lem}{Lemma} \newtheorem{thr}{Theorem}
 \newcommand{\edf}{\stackrel{\rm def}{=}} \def\l{\lambda}

 \title {Computational Complexity of Functions\thanks
 {{\em Theoretical Computer Science}  157:267-271, 1996.
 Partial translation from [Levin 74] (preliminary version is in [Levin 73]).}}

\author {Leonid A.~Levin\thanks {Supported by NSF grant CCR-9015276.}\\
Boston University\thanks {Computer Science department, 111 Cummington St,
Boston, MA 02215; (e-mail to Lnd@bu.edu).}} \date{} \maketitle\begin{abstract}

Below is a translation from my Russian paper.
 I added references, unavailable to me in Moscow.  Similar results have been
also given in [Schnorr Stumpf 75] (see also [Lynch 75]).  Earlier relevant work
(classical theorems like Compression, Speed-up, etc.) was done in [Tseitin 56,
Rabin 59, Hartmanis Stearns 65, Blum 67, Trakhtenbrot 67, Meyer Fischer 72].

I translated only the part with the statement of the results. Instead of the
proof part I appended a later (1979, unpublished) proof sketch of a slightly
tighter version. The improvement is based on the results of [Meyer Winklmann
78, Sipser 78]. Meyer and Winklmann extended earlier versions to machines with
a separate input and working tape, thus allowing complexities smaller than the
input length (down to its $\log$). Sipser showed the space-bounded Halting
Problem to require only additive constant overhead. The proof in the appendix
below employs both advances to extend the original proofs to machines with a
fixed alphabet and a separate input and working space. The extension has no
(even logarithmic) restrictions on complexity and no overhead (beyond an
additive constant). The sketch is very brief and a more detailed exposition is
expected later: [Seiferas Meyer].

 \end{abstract} \section*{Some Remarks}

We formulate the theorems in terms of the Turing Machine space. But it is clear
how to generalize them, since any complexity measure is bounded by a total
recursive function (t.r.f.) of any other one. Of course, the accuracy of a
constant factor will turn into the accuracy of some other t.r.f.\ We consider
one tape Turing Machines with arbitrary tape alphabets. If the alphabet has $n$
symbols, then input and output integers are written in the $n$-ary number
system. The {\em space} $p_A(x)$ of an algorithm $A$ is the size [reduced by 1]
of the tape used by $A(x)$. The length of a word $x$ is denoted $l(x)$.
Obviously, $p_A(x)+1 \ge \max(l(x),l(A(x)))$.

The space complexity of any function can be reduced by any constant
factor, by extending the alphabet. The inequality within a constant factor
$f\prec g$ means $\exists C\forall x\ f(x)\le Cg(x)$.

	Every function $F$ is associated with a class of algorithms that
compute it and with the class of their space complexities $M_F$. We
characterize all such classes extending well known Compression and Speed-up
Theorems. Some computable functions do not belong to any class $M_F$:

{\bf Note:} A partial function $p$ can be a space of an algorithm if and only
if it is itself computable within space $p(x)$. We call such functions {\em
simple}. This requirement is weak since usual functions $p$ are computable in
space $l(p(x))=\log p(x)$.

	We call {\em simple} an algorithm which outputs its own space. We
define $p_A(x)=\infty$ when $A(x)$ does not halt and interpret inequalities
with simple functions accordingly.
 Let us agree that an algorithm computes a function $F$, if it does this
everywhere in the intersection of its domain and the domain of $F$.

\newpage \section*{Formulation of the Theorems.}

For any {\em simple} function $G$, Compression Theorem [Rabin] provides a
function $F$, computable in exactly those spaces $p$ which are simple and
$p\succ G$. We generalize this theorem for an arbitrary recursive $G$:

\begin{thr} For any t.r.f.\ $G$ there exists a t.r.f.\ $F$,
with range $\{0,1\}$ computable in exactly those spaces $p$ which are
simple and $p\succ G$.\end{thr}

	Compression Theorem describes a very special case of t.r.f.\ In [Blum]
t.r.f.\ were discovered which have no such exact simple lower bounds of
complexity. However, the above generalization of the Compression Theorem
already describes the general case and can be inverted as follows:

\begin{thr} For any t.r.f.\ $F$ there exists a t.r.f.\ $G$
such that $F$ is computable in exactly those spaces $p$ which are simple
and $p\succ G$. \end{thr}

	Thus, the complexity class of any t.r.f.\ is organized naturally,
despite the Speed-up Theorem. The point is that the set of t.r.f.\ is richer
than the set of simple functions. Naturally, the complexity of an arbitrary
t.r.f.\ cannot be always characterized by a simple function, though it is
always characterizable by a t.r.f.\\
 Let us describe the properties of the complexity classes for arbitrary t.r.f.\
A class $M$ is called {\em canonical} if: \begin{enumerate}

 \item All functions of $M$ are simple, and some of them are total and

 \item If $f,g,h$ are simple, $f,g\in M$, and $h\succ\min(f,g)$, then
$h\in M$ and

 \item The class $\ov M$ of simple algorithms, computing the functions of
$M$, is of type $\Sigma^0_2$, i.e.\ can be defined as $(p\in\ov M)
\Longleftrightarrow\exists a\forall b\ R(a,b,p)$, where $R$ is recursive.
 \end{enumerate}

\begin{thr} $M$ is the class $M_F$ of all space complexities of some t.r.f.
$F$ iff M is canonical. \end{thr}

	 This theorem justifies the following conjecture of A.N.~Kolmogorov:
for any ``good" decreasing sequence of functions $p_i$ there exists a function,
computable with such and only such space complexities, that exceed some of the
$p_i$'s. The Compression and Speed-up Theorems are special cases. This
conjecture also describes the general case of complexity as it follows from
Lemma 1 below and Theorem 3. The above results extend to the case of partial
functions:

\begin{thr} Let $A$ be an r.e.\ set.  Theorems~1-3 remain valid, if the term
``t.r.f." is replaced everywhere by ``partial r.f.\ with domain $A$", and
inequalities like ``$a\succ b$" are restricted to $x\in A$. \end{thr}

\section*{Proofs.}	We call {\em $A$-canonical} a class $M$, satisfying
conditions~1-3, as adjusted in Theorem~4.

\begin{lem} For any $A$-canonical $M$, a p.r.f.\ $g$ exists, non-increasing
with $k$ and such that $g(k,x)+l(k)$ is simple, domain of $g(0,x)$ is
$A$ and $p\in\ov M\Longleftrightarrow(\exists k\,g(k,x) \prec p(x))$, for
any simple $p$.  \end{lem}

[$\ldots$]

\newpage \section* {Appendix (not part of the translation).}

Below is the sketch of a proof of a slightly tighter statement. It assumes
separate input and working space, thus allowing spaces $o(|x|)$, as in [Meyer
Winklmann 78]. It also assumes a fixed tape alphabet, allowing additive (rather
than multiplicative) constant accuracy. The latter uses the result of [Sipser
78] that the space $s$ bounded halting problem can be solved in space $s+O(1)$.
Otherwise, the version is similar to the above translation. For $\log$ of time
of a Pointer Machine or of some Turing Machine versions [Levin 91] similar
results hold.

\paragraph {Model.} To allow space limits below the input bit-length $|x|$ one
needs to differentiate the input symbols from symbols used as memory during the
computation. Instead of separating the input tape as in [Hartmanis Stearns 65],
I prefer to separate the ``ink''. While not essential, this preserves the
simple space-time geometry of the one-tape Turing Machine (TM). So, we separate
the state of each cell into a read-only {\em ink} and a read-write {\em pencil}
part. The ink part cannot be modified after the input is written and is ignored
for measuring space. The ink (but not necessarily pencil) string starts at the
left end of the tape after exactly one blank. The ink and pencil string and
their union form each a continuous segment without blanks. The alphabet is
fixed and has at least two symbols $\{0,1\}$ besides the blank. {\em Space}:
$S_{A(x)}$ or $S_A(x)$ is the supremum of bit-lengths of the pencil string
throughout the computation of $A(x)$. The output may either be left on the
tape/head or its digits ``flashed'' sequentially at the (fixed) leftmost cell.
In some cases the pencil string starts not empty. E.g.\ $g$-constructible
functions $f$ are those computable in space $\max\{t,f(x)\}$ starting from
input $x$ and any pencil string of length $t\ge g(x)$; for $g=0$ we omit the
prefix ``$g$-'' and for $g=f$ replace it with ``semi-''.

 \paragraph {Conventions.} Let $U(k,x)$ be a Universal TM with $\{0, 1\}$
outputs. It ignores the ``padding'' $k_2$ in its program $k=(k_1,k_2)$.
Appropriate paddings can put any $\Sigma^0_2$ program set in the form
$m=a^{-1}(\{\infty\})$ for some function $a$ with constructible $a(k)-4|k|$.
 Let $p(k,x)\edf4|k|+ S_U(k,x)$ and expressions like $p_k(x)$ mean $p(k,x)$.
 Assume $0\in m$ and $M=\{p_k:k\in m\}$. Consider the closure $\ov M$ of a set
$M$ of functions under inclusion of each $h$ s.t.\ for some $f,g\in M$,
$h\ge\min\{f,g\}-1$ in the domain $D$ of $U(0,x)$. Call sets $M_1,M_2$
 {\em cofinal} if their closures contain the same constructible functions.
Define $[a<b]$ as $a$, if $a<b$ and $0$ otherwise. Likewise for $\le$. Clearly,
the complexity class of any function can be described as $\ov M$ above.

\paragraph{Construction.} Now we build (cf.\ Lemma 1) a monotone sequence $g_k$
cofinal to $M$:\\ If $a(k)>t>p(k,x)$, let $p^t(k,x)\edf
\max_{l<k}\{p(k,x),[a(l)\le t]\}\le t$. Otherwise, $p^t(k,x)\edf t$.\\
 Then, $g(1,x)\edf p(0,x)$; \ \ $g(k\!+\!1,x)\edf p^{g(k,x)}(k,x)$; \ \
$g_\infty(x)\edf\min_k g(k,x)$; \ \ $k_x\edf\min\{l:g(l,x)= g(k,x)\}$.\\
 To compute $g_k(x)$ we carry $k$, $g_{k-1}(x)$ as the pencil string length,
and the largest relevant $a(l)$ as $g_{k-1}(x)-a(l)$ (if $<2|k|$) or as $l$.
 Cutting the values of $p,a$ to the maximum of $t$ would not affect those
values of $g$ below $t$. So, $g(k,x)-2|k_x|$ is $g_\infty$-constructible;
$g_k$ are uniformly recursive with domain $D$ and equal $\min_{l<k}\{p_l(x):l\!
\in\!m\}$, on $D$, except when both are $\le\max_{l<k}\{[a(l)<\infty]\}=O(1)$.

Next we convert such $\{g_k\}$ into a cofinal set consisting of a single
semiconstructible recursive function $G$ on $D$ (cf.\ Theorem 2): Let $b(k)\edf
\min_x(2|k,x|+g_1(x):g_k(x)>p_k(x))$ and $K(x)\edf \min\{k: b(k)>g_k(x)>
p_k(x)\}$.  Then $G(x)\edf g(K(x),x)\le\max_{l\le k}\{g(k,x),[b(l)<\infty]\}$,
for all $k$.\\ Conversely, $G<p_k$ in $D$ implies $b(k)=\infty$. Indeed, $b(k)=
2|k,x|+g_1(x)<\infty$ while $g_k(x)>p_k(x)$ yields $K(x)\le k$ and $G(x)\ge g_k
(x)>p_k(x)$. $b(k)=\infty$ makes $g_k(x)\le p_k(x)$ in $D$ and $p_k\in\ov M$.

Finally, for such $G$, we build a $G$-constructible predicate $\Pi(x)\edf
1-U(K'(x),x)$ with complexity class cofinal to $G$ (cf.\ Theorem 1):\\
 Here $K'(x)\edf\min\{k:c(k)>G(x)\ge p_k(x)\}$ and $c(k)\edf
 \min_x\{2|k,x|+p_0(x): K'(x)=k\}$.\footnote {A leaner version:
 $c(k)\edf\min_x\{2|k,x|+\max\{p_0(x),p_k(x)\}:\Pi(x)\ne U(k,x)\}$.}
 If $\Pi(x)=U(k,x)$ in $D$ then $k\not\in K'(D)$ and $c(k)=\infty$.
 Then $G(x)\le\max_{l<k}\{p_k(x),[c(l)<\infty]\}$ and $p_k\in\ov M$.

\newpage\subsection{Acknowledgments}

My work on this extension, as well as my smooth implantation into the Western
scientific community was made possible by comprehensive support and
encouragement by Albert Meyer in 1978-80. I am among the many greatly indebted
to his concern and generosity.

  \end{document}